\documentclass{article}
\usepackage[T1]{fontenc} 
\usepackage[utf8]{inputenc} 
\usepackage{ismir,amsmath,cite,url}
\usepackage{graphicx}
\usepackage{color}
\usepackage{caption}
\usepackage{subcaption}
\usepackage{siunitx}
\usepackage{booktabs}
\usepackage{enumitem}
\usepackage{stackengine}
\usepackage{amsfonts}

\title{Musical Tempo Estimation Using a Multi-scale Network}

\multauthor
{Xiaoheng Sun$^1$ \hspace{1cm} Qiqi He$^1$ \hspace{1cm} Yongwei Gao$^1$  \hspace{1cm} Wei Li$^{1,2}$} {
 $^1$  School of Computer Science and Technology, Fudan University, Shanghai, China\\
$^2$ Shanghai Key Laboratory of Intelligent Information Processing, Fudan University, Shanghai, China\\
{\tt\small \{19210240112, heqq20, ywgao16, weili-fudan\}@fudan.edu.cn}}

\def\authorname{Xiaoheng Sun, Qiqi He, Yongwei Gao and Wei Li}

\begin{document}

\maketitle
\begin{abstract}
Recently, some single-step systems without onset detection have shown their effectiveness in automatic musical tempo estimation. Following the success of these systems, in this paper we propose a Multi-scale Grouped Attention Network to further explore the potential of such methods. A multi-scale structure is introduced as the overall network architecture where information from different scales is aggregated to strengthen contextual feature learning. Furthermore, we propose a Grouped Attention Module as the key component of the network. The proposed module separates the input feature into several groups along the frequency axis, which makes it capable of capturing long-range dependencies from different frequency positions on the spectrogram. In comparison experiments, the results on public datasets show that the proposed model outperforms existing state-of-the-art methods on Accuracy1.
\end{abstract}

\section{Introduction}\label{sec:introduction}

Although there are many different ways to describe musical tempo (e.g., measures per minute, bars per minute, or even a range of Italian terms), beats per minute (BPM) is the most commonly used measurement unit. The estimation of BPM plays an important role in a variety of applications, such as music recommendation, automatic accompaniment, playlist generation, etc. Because of its utility, the automatic estimation of tempo has been an important task and received continuous attention in the field of music information retrieval (MIR) \cite{goto1994beat, scheirer1998tempo, gouyon2006experimental, schreiber2020data}. 

Traditional methods for automatic tempo estimation are usually based on hand-crafting signal processing. To estimate the tempo of a given audio segment, an onset strength signal (OSS) function is firstly derived, and the frequency of the major pulses is extracted and converted to BPM. The OSS function is a function whose peaks should correspond to onset times. It can be obtained by various methods, such as means of auto-correlation \cite{dixon2001automatic, alonso2006accurate}, comb filters \cite{scheirer1998tempo, klapuri2005analysis} and Fourier analysis \cite{cemgil2000tempo}. Machine learning techniques are also adopted for tempo estimation, including Gaussian mixture models (GMM) \cite{peeters2012perceptual}, support vector machines (SVM) \cite{gkiokas2012reducing, percival2014streamlined}, k-nearest neighbors (k-NN) \cite{wu2014supervised, wu2015musical}, random forests \cite{schreiber2017post} and so on. Since B{\"o}ck \cite{bock2015accurate} proposed a recurrent neural network (RNN) model to learn beat-level representations from audio signals, attempts to use deep neural networks (DNN) for tempo estimation began to grow \cite{gkiokas2017convolutional, bock2019multi, bock2020deconstruct}. 

In all methods mentioned above, the extraction of BPM depends on some post-processing of OSS functions or beat activation functions. It is only in recent years that the \emph{single-step} tempo estimation systems based on DNN appeared. As the first single-step approach for tempo estimation, the CNN model proposed by Schreiber \cite{schreiber2018single} is capable of extracting BPM value directly from a Mel-scaled spectrogram. In this work, classification is proved to be an effective method for tempo estimation. Adopting a similar idea, Foroughmand \cite{foroughmand2019deep} proposed the Harmonic-Constant-Q-Modulation (HCQM), a new representation of audio signal, as the input of a relatively simple CNN classification model. The experimental results also showed its effectiveness.


A commonly used metric in tempo estimation is Accuracy1 \cite{gouyon2006experimental}, indicating the percentage of correct estimates allowing a $\pm4\%$ tolerance. However, automatic tempo estimation systems tend to predict a wrong tempo by a factor of 2 or 3, known as \emph{octave errors}. As an additional measure, Accuracy2 is introduced, which ignores octave errors. In some applicational scenarios (such as DJ software), accurate tempo annotations are mandatory and octave errors are unacceptable \cite{gartner2013tempo}, but most existing algorithms' performance on Accuracy1 is still far from satisfactory.

Previous works \cite{schreiber2018single, foroughmand2019deep} have shown the potential of CNN-based single-step approach to improve performance on Accuracy1. Following the success of these methods, in this paper we propose a CNN-based single-step model named Multi-scale Grouped Attention Network (MGANet). A multi-scale network architecture is designed to aggregate information from different scales to produce superior feature representations. Furthermore, a Grouped Attention Module (GAModule) is proposed to capture long-range dependencies and refine the feature based on the attention mechanism. 

The remainder of this paper is organized as follows. In Section 2, we introduce the proposed method in detail. In Section 3, experimental results are presented to show the effectiveness of our method. Finally, we make further conclusion in Section 4.

\begin{figure*}[t]
    \centering
    \includegraphics[width=17.2cm, trim={2.8cm 8.7cm 4cm 5.5cm},clip]{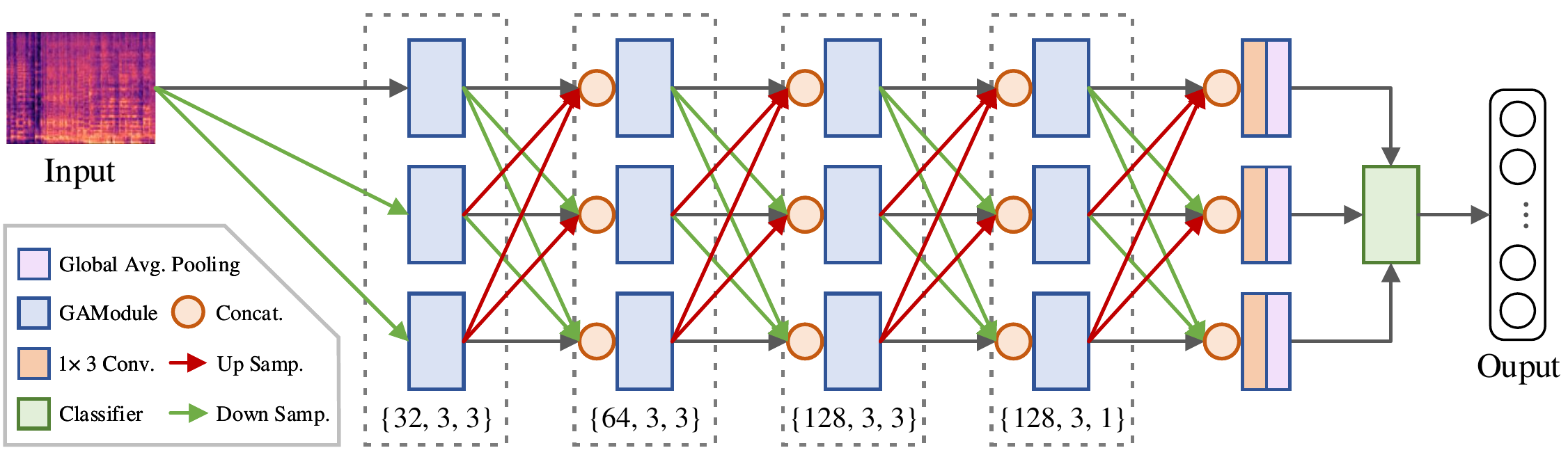}
    \caption{The overall architecture of Multi-scale Grouped Attention Network (MGANet). The numbers in dashed boxes indicate the three parameters of GAModules: \{output channel number $C$, pooling size $p$, group number $k$\}. Every concatenation operation in the figure is followed by a $1\times$1 convolution layer to adjust channel number. The classifier consists of a concatenation operation, a fully connected layer, and a softmax layer.}
    \label{fig:network}
\end{figure*}

\section{Approach}

\subsection{Proposed Model}
Same as \cite{schreiber2018single} and \cite{foroughmand2019deep}, we also treat tempo estimation as a classification problem. The output of our model is a probability distribution of 256 BPM classes (from 30 to 285 BPM). Because the Mel-scaled frequency matches closely the human auditory perception, we choose the Mel-scaled spectrogram as the raw feature. First, the original audio data is resampled to 11.025 kHz. Then, we use half-overlapping windows of 1,024 frames, and transform each window into an 81-band Mel-scaled magnitude spectrum. The input of the proposed model is designed as a spectrogram segment of 128 frames, roughly 6 seconds long.

In the rest of this section, we first present the overall architecture of the proposed MGANet. Then, we introduce the GAModule, which is the key component of the network.

\subsubsection{Multi-scale Network Architecture}
The goal of tempo estimation is to extract a periodic pattern from an audio signal. Therefore, global information of the input spectrogram is particularly important. Due to the characteristics of CNN, overall pattern extraction is usually achieved by stacking multiple layers. 
But directly repeating convolution layers makes the model difficult to design and optimize. Another way is to use large-size convolution kernels to enlarge the receptive fields. 
However, this is also costly because of the increase in parameters and multiply-add operations. 
To solve the problem, we introduce the idea of multi-scale structure, which has been proved to be effective in many classification tasks \cite{huang2018multi, wang2020deep, adegun2020fcn}. By downsampling / upsampling the feature to different scales and exchanging information repeatedly, high-level representations can be derived after just a few layers. 


As shown in Figure \ref{fig:network}, the overall architecture of MGANet is mainly composed of three branches for different scale. 
In each branch, input features are gradually downsampled over the frequency (vertical) axis, but maintains the resolution through the whole process on the time (lateral) axis. Furthermore, these feature maps from different scales are merged repeatedly to integrate contextual information, leading to high-level representations amenable to classification. 

Specifically, the input spectrogram is first downsampled by 1/2 and 1/4 over the time axis with average pooling, resulting in three representations of sizes (81, 128), (81, 64), and (81, 32). Then, the representations are fed into three parallel branches respectively to perform feature processing. The processing is mainly done by the proposed GAModule described in section 2.1.2. 
Through the whole structure, we repeat multi-scale fusion by rescaling and concatenation. Average pooling and transposed convolution \cite{xiao2018simple} layers with kernel size of $1\times3$ are used to perform rescaling. For concatenation, a $1\times1$ convolution layer with the exponential linear unit (ELU) \cite{clevert2015fast} activation is followed to adjust the channel number.

Processed by GAModules, the features are gradually downsampled over the frequency axis to summarize frequency bands, making the representations easier to detect periodicity. On each branch, the downsampling is repeated four times. Accordingly, the channel numbers of the features are increased. After the above processes, three feature maps with shapes (1, 128, 128), (1, 64, 128), and (1, 32, 128) are obtained. Then, these feature maps are fused again and fed into a $1\times3$ convolution layer to adjust channel numbers to 256. After global average pooling, three vectors of length 256 are concatenated together. Finally, a fully connected layer takes the vector as input and a softmax layer is used to derive the probability distribution of 256 tempo classes.

\begin{figure}[t]
    \centering
    \includegraphics[width=7.0cm, trim={4.5cm 2.4cm 14.9cm 3.8cm}, clip]{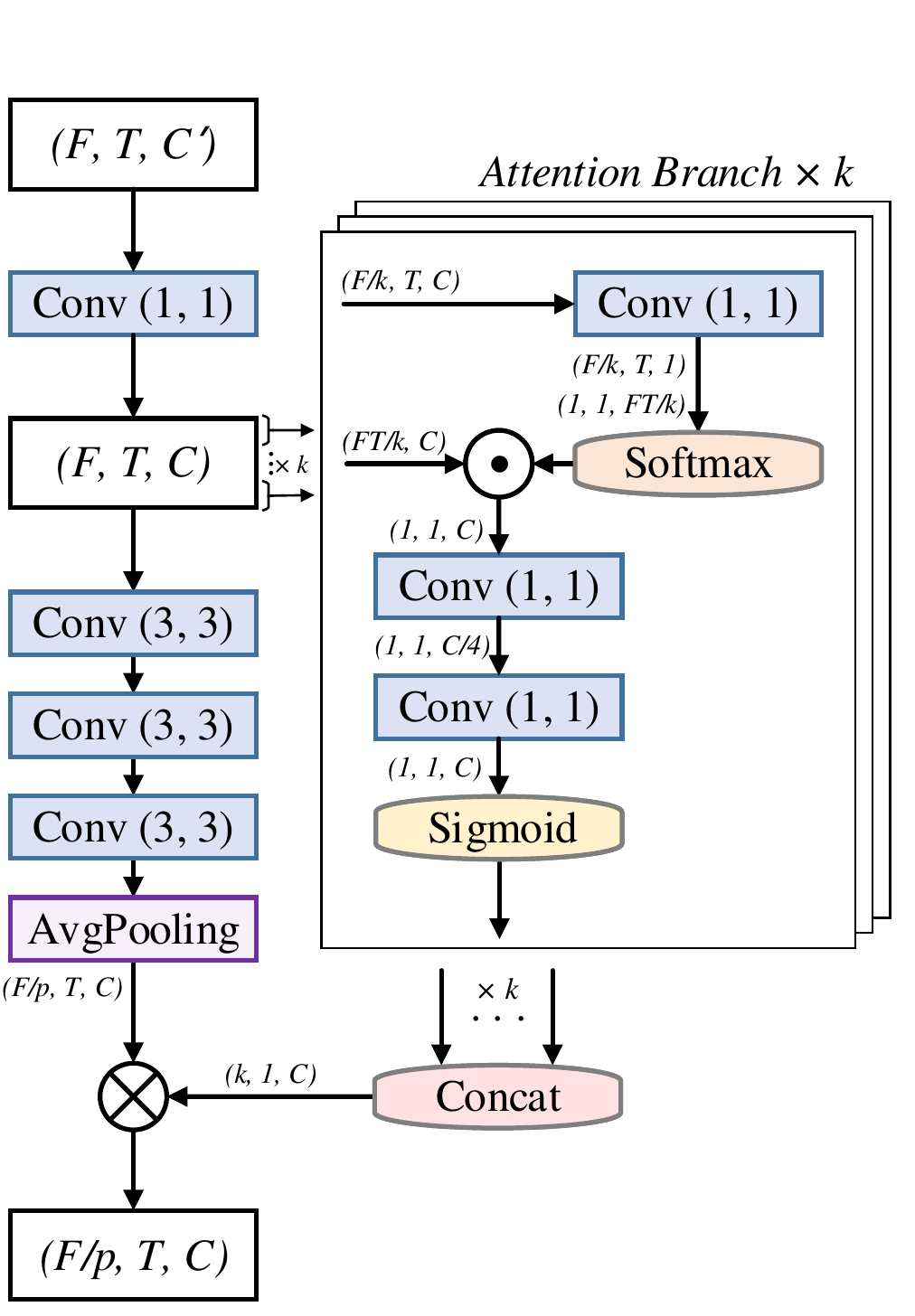}
    \caption{The structure of Grouped Attention Module (GAModule). Feature maps are shown as feature dimensions, e.g. $(F, T, C)$ denotes a feature map with height $F$, width $T$, and channel number ${C}$. $p$ and $k$ denote pooling size and group number respectively. ${\odot}$ denotes matrix multiplication and ${\otimes}$ denotes broadcast element-wise multiplication.}
    \label{fig:module}
\end{figure}

\subsubsection{Grouped Attention Module}

The proposed GAModule structure is shown in Figure \ref{fig:module}. The module consists of two parts: a trunk branch performing feature processing, and $k$ attention branches producing an attention mask to capture global context information and recalibrate the output feature map. 

The structure of the attention branch is mainly inspired by the global context network (GCNet) \cite{cao2019gcnet}, which is designed for long-range dependency modeling through attention mechanism. The attention mechanism biases the allocation of the most informative feature expressions and suppresses the less useful ones. Recently, the benefits of the attention mechanism have been demonstrated in a series of tasks. 
We introduce the attention mechanism into GAModule mainly for two purposes: 1) model the long-range dependencies to obtain global context features; 2) reweight the importance of different channels to improve the representational capacity of the refined feature. 

Unlike the images in the field of computer vision, the two axes of audio spectrograms have different meanings, which respectively represent frequency and time. Furthermore, it is known that different musical instruments have different frequency ranges, and different frequency ranges have a different impact on the total sound. These facts indicate that different frequency bands contain relatively independent information. 
Based on these observations, we believe that it's inappropriate to aggregate the whole spatial scope at once to calculate long-range dependencies. Instead, different frequency positions of the feature should be handled separately, which will help to filter the useful information more efficiently. Therefore, different from traditional channel-wise attention models that aggregate the entire feature to generate one attention map (e.g., squeeze-and-excitation networks \cite{hu2018squeeze}), we divide the feature equally into $k$ groups along the frequency axis and send each fragment into an independent attention branch. We termed the operation as \emph{grouped channel attention}. 


As shown in Figure \ref{fig:module}, the framework of the attention branch is roughly the same as the GC block in GCNet. Firstly, the feature map is squeezed into a channel descriptor by global attention pooling. The pooling is achieved by convolution, softmax, and matrix multiplication. For an input feature map $x$, the generated descriptor $z\in\mathbb{R}^C$ is calculated by

\begin{eqnarray}
z=\sum\nolimits_{j=1}^{N_p}\frac{exp(ELU(Wx_j))}{\sum\nolimits_{m=1}^{N_p}exp(ELU(Wx_m))}x_j
\label{eq:eq1}
\end{eqnarray}

\noindent where $j$ and $m$ enumerate all possible positions, and $W$ denotes linear transformation matrix. We adopt ELU as the activation of the convolution layer to further increase robustness. After the pooling, global spatial information is gathered in the descriptor. Then, a bottleneck of two-layer architecture is formed to transform information. We adopt a reduction ratio of 4 and ELU activation in the first layer. A sigmoid function is then applied to rescale the transformation output. Finally, $k$ attention maps with the shape of $(1, 1, C)$ can be obtained. We concatenate these attention maps along the frequency axis and get the output attention map of $(k, 1, C)$.

Simultaneously, in the trunk branch we simply stack three convolution layers with kernel of $3\times3$ and ELU activation. Because of the existence of attention branches, the trunk does not need a complex structure and too many layers, which reduces the number of parameters and the complexity of the model. We use average pooling with pooling size of $p\times1$ to downsample the feature map to $(F/p, T, C)$. Finally, broadcast element-wise multiplication is performed to fuse the output of the trunk branch and attention branches. Through the fusion, the output feature map is refined by global contextual information gathered by grouped attention operation.

\subsection{Training Data \& Augmentation}

For training and validation, we adopt the three training datasets used in \cite{schreiber2018single}: \emph{LMD Tempo} (3,611 items), \emph{MTG Tempo} (1,159 items), and \emph{Extended Ballroom} (3,826 items). However, though covering multiple musical genres, the combination of these datasets is not genre-balanced, and some common genres are even missing. It is known that tempo perception is closely related to music genre. 
For example, for popular music, people usually perceive tempo through drumbeats, while for classical music, people often perceive tempo from bass instruments such as double bass. 
To alleviate the genre imbalance, we use two additional datasets to supplement the training data:

\begin{itemize}[leftmargin=*]




\item \textbf{\textit{RWC-popular}}: To further enhance the model's ability to estimate pop music tempo, we used RWC-popular \cite{goto2002rwc} (a pop music database with 100 pieces) for training. We cut the songs into 30s fragments without overlapping, resulting in 735 items.

\item \textbf{\textit{FD-Tempo}}: To enrich the genres of training data, we selected some tracks of classical music. For each track, we chose several 30s excerpts with stable tempi and annotated them by manually tagging. Finally, 530 items are obtained as an additional dataset termed \textit{FD-Tempo}.

\end{itemize}

We use the combination of the five datasets for training and validation. It contains 9,861 tracks with a total length of 41h 3min. Specifically, we randomly choose 500 tracks for validation, and the rest 9,361 tracks are used for training.


\begin{figure}
    \centering
    \includegraphics[width=7.5cm]{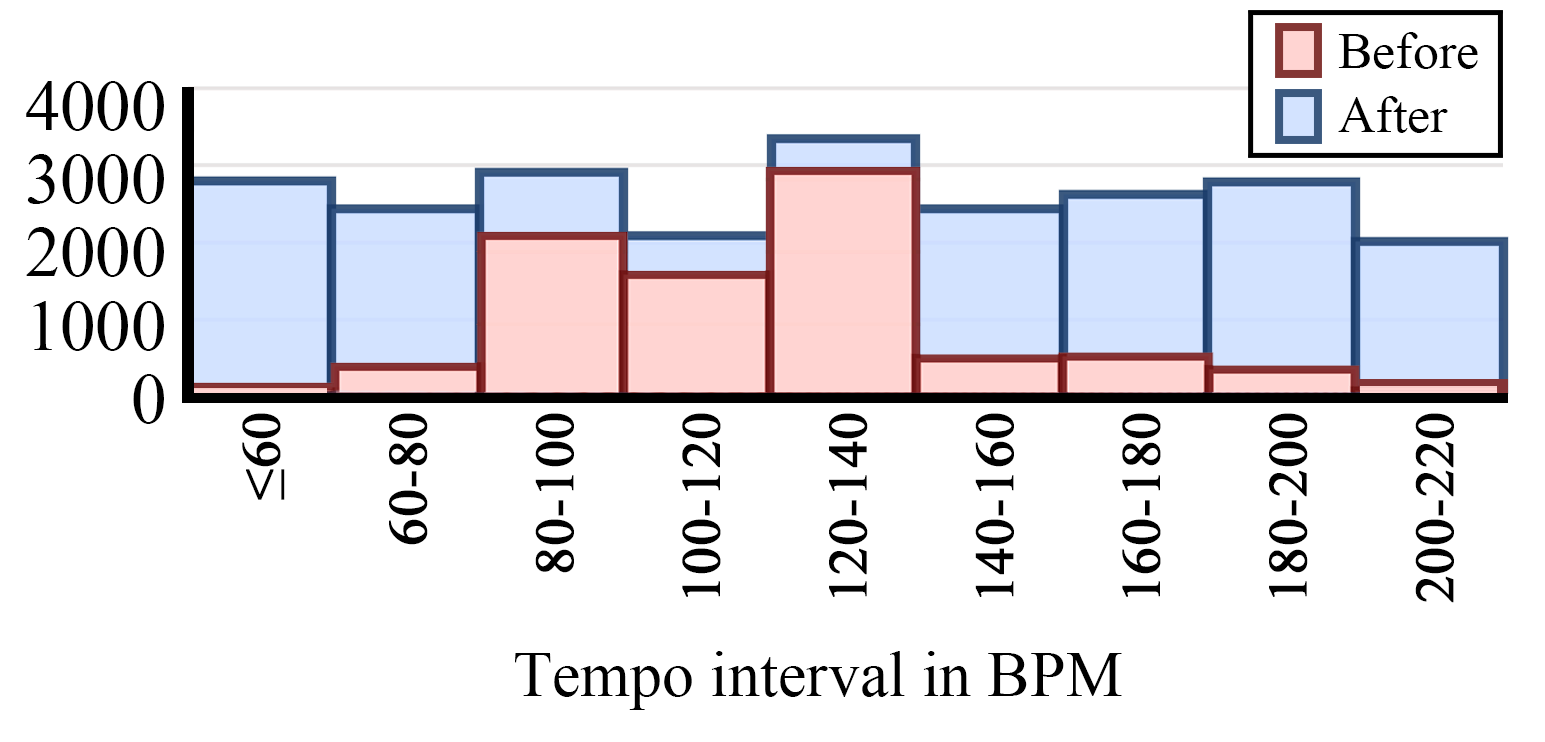}
    \vspace{-0.4cm}
    \caption{Tempo distribution before and after augmentation.}
    \label{fig:tempoDistribution}
\end{figure}

To alleviate the BPM class imbalance, we further augment the training set by speeding up / slowing down the selected tracks with factors randomly chosen from 0.7$\sim$1.4 without altering the pitch. 
We retain the original files and make sure that the same audio will not be selected more than 15 times. 
After augmentation, the number of tracks increases from 9,361 to 23,512. Note that the validation set is not augmented. The tempo distribution in the training set before and after augmentation is shown in Figure \ref{fig:tempoDistribution}. Besides, we also adopt the scale-\&-crop data augmentation mentioned in \cite{schreiber2018single} to further increase the variability of training data.

\subsection{Training Details}

For training, the batch size we set is 32. In each epoch, 128 consecutive frames of each sample are randomly selected for training. We choose the categorical cross-entropy as the loss function, and an Adam optimizer \cite{kingma2014adam} is applied with a learning rate of 0.001. We evaluate Accuracy1 of the validation set every 500 iterations, and save the model with the highest accuracy. The training is not stopped until Accuracy1 has not improved for 50,000 iterations.

\section{Evaluation}

We choose Accuracy1 (ACC1) and Accuracy2 (ACC2) \cite{gouyon2006experimental} as the evaluation metrics. Accuracy1 is defined as the percentage of correct estimates allowing a $\pm4\%$ tolerance. Accuracy2 ignores octave errors by a factor of 2 and 3, and also allows a $\pm4\%$ tolerance. As mentioned earlier, the demand for highly accurate tempo annotations has become increasingly urgent in many applicational scenarios. Hence we mainly focus on improving Accuracy1.

We focus on the performance on global tempo estimation based on the assumption the tempo of the input track stays constant, and only one BPM value will be returned by the estimation system. In the experiment, the global tempo is obtained by averaging the outputs of softmax layer over different parts of a full track \cite{schreiber2018single}. 

\subsection{Ablation Study}

\begin{table}[t]
    
    \begin{center}
    \vspace{-0.2cm}
    
    \begin{subtable}[c]{0.20\textwidth}
        \centering
        
        \begin{tabular}{l|cc}
        \toprule[1pt]
        Method&ACC1&ACC2\\ \hline\hline
        w/o AB & 77.0 & 89.9\\
        w/o GA & 78.5 & 89.1\\
        Single-scale & 75.8 & 89.6\\\hline
        Proposed & \textbf{78.9} & \textbf{91.3}\\\bottomrule
        \end{tabular}
        \subcaption{GTzan}
       \label{tab:GTzan}
    \end{subtable}
    \hfill
    \begin{subtable}[c]{0.20\textwidth}
        \setlength{\tabcolsep}{3mm}{
        \centering
        
        \begin{tabular}{cc}
        \toprule[1pt]
       ACC1&ACC2\\ \hline\hline
       79.8 & 95.3\\
       79.0 & 94.2\\
       71.2 & 94.5\\\hline
       \textbf{82.1} & \textbf{95.7}\\\bottomrule
        \end{tabular}
        \subcaption{ACM Mirum}
       \label{tab:ACM Mirum}
       }
    \end{subtable}
    \caption{Results of ablation study. "w/o AB" and "w/o GA" denote "without attention branch" and "without grouped attention" respectively. 
    Best results are set in \textbf{bold}.}
    \label{tab:compTable}
    \end{center}
    
    \vspace{-0.5cm}
    
\end{table}

We study the effect of each idea in our approach. To simplify the discussion, we select two test datasets \emph{GTzan} \cite{marchand2015gtzan} and \emph{ACM Mirum} \cite{peeters2012perceptual} for analysis. These two datasets are relatively large (999 and 1,410 items respectively), and both cover rich genres. 

To investigate how much the proposed GAModule contributes to the model, we design a set of experiments. Firstly, we remove the attention branches in the module, and only the trunk branch is remained to process features. As shown in Table \ref{tab:compTable}, the performance degrades for both datasets. When focusing on Accuracy1, the performance decreases by 1.9\% for \emph{GTzan} and 2.3\% for \emph{ACM Mirum}. Then, in another experiment we keep only one attention branch in each module, which can be achieved by setting GAModules' parameter $k$ to 1. The Accuracy1 reduced by 0.4\% and 3.1\% respectively. For Accuracy2, in both experiments there is also a certain degree of decline. These results indicate that the attention mechanism is helpful to capturing long-range dependencies and therefore improve the generalization of the model. But directly using existing modules may hinder the effect. The proposed grouped attention takes into account the characteristics of spectrogram and achieves further improvements of the model.

Then, we analyze the effect of the multi-scale architecture by changing the architecture to a single-scale one. We remove all downsampled subnetworks and only retain the one with the highest resolution (the topmost branch in Figure \ref{fig:network}). As shown in Table \ref{tab:compTable}, model without multi-scale architecture shows significantly worse performance on Accuracy1. The Accuracy1 decreases by 3.1\% and 10.9\% for \emph{GTzan} and \emph{ACM Mirum} respectively. For Accuracy2, there is also a certain degree of performance degradation. The results demonstrate that the multi-scale can improve the classification ability as well as robustness.

\subsection{Comparison with Previous Work}

\begin{table*}[t]
    \resizebox{\textwidth}{24.8mm}{
    \begin{subtable}[c]{0.45\textwidth}
        \centering
        \begin{tabular}{l|S[table-format=2.1, table-space-text-post = {$^{*}$}]cc|c}
        \toprule[1pt]
        \textbf{Dataset}&\texttt{b\"{o}ck}&\texttt{schr}& \texttt{foro} & \texttt{mgan} \\ \hline\hline
        ACM Mirum          &74.0                &79.5   &73.3   &\textbf{82.1}\\
        Hainsworth         &\textbf{80.6}$^{*}$ &77.0   &73.4   &77.5\\
        GTzan              &69.7                &69.4   &69.7   &\textbf{78.9}\\
        SMC                &\textbf{44.7}$^{*}$ &33.6   &30.9   &29.0\\
        GiantSteps         &58.9                &73.0   &83.6   &\textbf{90.2}\\
        Ballroom           &84.0$^{*}$          &92.0   &92.6   &\textbf{95.1}\\
        ISMIR04            &55.0                &60.6   &61.2   &\textbf{61.7}\\\hline
        Combined           &69.5                &74.2   &74.4   &\textbf{79.8}\\\bottomrule
        \end{tabular}
       \subcaption{Accuracy1}
       \label{Accuracy1}
    \end{subtable}
    
    \hfill
    \begin{subtable}[c]{0.45\textwidth}
        
        \centering
        \begin{tabular}{l|S[table-format=2.1, table-space-text-post = {$^{*}$}]cc|c}
        \toprule[1pt]
            \textbf{Dataset}&\texttt{b\"{o}ck}&\texttt{schr}& \texttt{foro} & \texttt{mgan} \\ \hline\hline
        ACM Mirum          &\textbf{97.7}       &97.4   &96.5       &95.7\\
        Hainsworth         &\textbf{89.2}$^{*}$ &84.2   &82.9       &87.8\\
        GTzan              &\textbf{95.0}       &92.6   &89.1       &91.3\\
        SMC                &\textbf{67.3}$^{*}$ &50.2   &50.7       &44.7\\
        GiantSteps         &86.4                &89.3   &\textbf{97.9}&97.6\\
        Ballroom           &\textbf{98.7}$^{*}$ &98.4   &\textbf{98.7}&97.7\\
        ISMIR04            &\textbf{95.0}       &92.2   &87.1       &88.8\\\hline
        Combined           &\textbf{93.6}       &92.1   &92.0       &91.9\\ \bottomrule
        \end{tabular}
        \subcaption{Accuracy2}
        \label{Accuracy2}
    \end{subtable}
    }
    \caption{Comparison with the results published by B\"{o}ck (\texttt{b\"{o}ck}) \cite{bock2015accurate}, Schreiber (\texttt{schr}) \cite{schreiber2018single}, and Foroughmand (\texttt{foro}) \cite{foroughmand2019deep}. 
    Best results per test dataset are set in \textbf{bold}. 
    Asterisk (*) denotes that the corresponding dataset were used for training.
    }
    \label{tab:comp}
\end{table*}

To compare with previous work, we use the same test datasets as in \cite{schreiber2018single} (see \cite{schreiber2017post} for details): \emph{ACM Mirum} \cite{peeters2012perceptual} (1,410 items), \emph{Hainsworth} \cite{hainsworth2003techniques} (222 items), \emph{GTzan} \cite{marchand2015gtzan} (999 items), \emph{SMC} \cite{holzapfel2012selective} (217 items), \emph{GiantSteps} \cite{knees2015two} (664 items), \emph{Ballroom} \cite{gouyon2006experimental} (698 items), and \emph{ISMIR04} \cite{gouyon2006experimental} (465 items). The union of all test datasets is referred to as \emph{Combined}. The most recent annotations available are used. 

We compare our work (\texttt{mgan}) with previous studies by Schreiber (\texttt{schr}) \cite{schreiber2018single} and Foroughmand (\texttt{foro}) \cite{foroughmand2019deep}. These two methods are both CNN-based single-step models that we are committed to improve. We consider them as the state-of-the-art among single-step approaches. In addition, we also compare the model with an RNN-based traditional periodicity analysis approach by B\"{o}ck (\texttt{b\"{o}ck}) \cite{bock2015accurate}. The results are shown in Table \ref{tab:comp}. Note that \emph{Ballroom}, \emph{Hainsworth}, and \emph{SMC} are used for training in \texttt{b\"{o}ck} (values marked with asterisks *).

Focusing on Accuracy1, the experimental results show that the proposed model surpasses other methods in most cases, which proves the effectiveness of the proposed idea to improve Accuracy1. 
Especially for \emph{GaintSteps} (664 electronic dance music excerpts), there shows a significant improvement of over 6.6\%. The richness of electronic dance music in training data can be considered as a reason. The good performance in \emph{ACM Mirum} and \emph{GTzan} (both multi-genre datasets) shows the generalization potential of our model. Moreover, for \emph{Hainsworth}, the model achieves the highest Accuracy1 among single-step approaches. Finally, the proposed method also reaches the highest Accuracy1 for \emph{Combined} (79.8\%) compared with other methods, gaining improvement of 5.4\%.

As for Accuracy2, it can be observed that \texttt{b\"{o}ck} achieves the highest accuracy in most cases. Ignoring \texttt{b\"{o}ck}, the proposed model shows a similar performance to other single-step methods. 

Among all datasets, the worst results of our model are obtained for \emph{SMC}. The dataset was designed to be difficult to estimate tempo, covering various genres. Although we have tried to supplement and augment the training data, the genre-imbalance problem has not been solved very well. This indicates the necessity to supplement more data with different genres in the future work. 


\subsection{Comparison with Multi-task Approaches}

\begin{table}[t]
        \centering
        \setlength\tabcolsep{13pt}
        \begin{tabular}{lcc}
        \toprule[1pt]
        & Accuracy1 & Accuracy2\\\hline
        
        \multicolumn{3}{c}{\emph{ACM Mirum}}\\
        \texttt{b\"{o}ck19} \cite{bock2019multi} & 0.749 & 0.974\\
        \texttt{b\"{o}ck20} \cite{bock2020deconstruct} & 0.841 & \textbf{0.990}\\
        \texttt{mgan} & 0.821 & 0.957\\
        \texttt{mgan+} & \textbf{0.846} & 0.970\\
        \multicolumn{3}{c}{\emph{GiantSteps}}\\
        \texttt{b\"{o}ck19} \cite{bock2019multi} & 0.764 & 0.958\\
        \texttt{b\"{o}ck20} \cite{bock2020deconstruct} & 0.870 & 0.965\\
        \texttt{mgan} & \textbf{0.902} & \textbf{0.976}\\
        \texttt{mgan+} & 0.861 & 0.973\\
        
        \multicolumn{3}{c}{\emph{GTzan}}\\
        \texttt{b\"{o}ck19} \cite{bock2019multi} & 0.673 & 0.938\\
        \texttt{b\"{o}ck20} \cite{bock2020deconstruct} & \textbf{0.830} & \textbf{0.950}\\
        \texttt{mgan} & 0.789 & 0.913\\
        \texttt{mgan+} & 0.796 & 0.931\\\bottomrule
        \end{tabular}

    \caption{Comparison with multi-task approaches. \texttt{mgan+} is trained by multi-task learning with beat tracking. Best results per test dataset are set in \textbf{bold}. }
    \label{tab:compMulti}
\end{table}

In recent years, some works \cite{bock2019multi, bock2020deconstruct} have not only focused on a single rhythm attribute, but combined the estimation of interconnected rhythm attributes (such as beats, downbeats, etc.) by multi-task learning, so that these highly related tasks can reinforce each other. These approaches are capable of embedding more musical knowledge into a single model, and enrich the training data of each task. In order to further explore the potential of the proposed MGANet and compare its performance with multi-task approaches, we conduct experiments with reference to \cite{bock2019multi}, combining the beat tracking task to our model. 

To predict beat positions, we add a branch to the original network structure. The inputs of the branch are the feature maps before sent into tempo classifier, with shapes of (1, 128, 128), (1, 64, 128), and (1, 32, 128). The low resolution feature maps are up-sampled to 128 frames length on time axis by transposed convolution layers. Then, the concatenated feature map with shape (1, 128, 384) is processed by three $1\times3$ convolution layers (output channel number are set to 128, 32, and 1 respectively). After a sigmoid operation, the beat activation function is derived. This extended network structure is trained as a multi-output model to combine the two tasks.

For the training of beat tracking, we use a combination of the following datasets: \emph{Hainsworth} \cite{hainsworth2003techniques}, \emph{SMC} \cite{holzapfel2012selective}, \emph{Ballroom} \cite{gouyon2006experimental}, \emph{ISMIR04} \cite{gouyon2006experimental}, \emph{Beatles} \cite{davies2009evaluation}, and \emph{HJDB} \cite{hockman2012one}. 
As for the training of tempo estimation, the training and validation datasets in section 2.2 are used. To further enrich the data, beat annotated datasets are also adopted for the training of tempo classifier, using the average BPMs derived from beat annotations as training labels. We train the two task alternatively every epoch, without changing other experimental settings mentioned in section 2.3.

The experimental results are shown in Table \ref{tab:compMulti}. Three datasets \emph{ACM Mirum} \cite{peeters2012perceptual}, \emph{GTzan} \cite{marchand2015gtzan}, and \emph{GiantSteps} \cite{knees2015two} are used as test datasets. We compare our works (the original model \texttt{mgan} and the multi-task model \texttt{mgan+}) with two multi-task approaches \texttt{b\"{o}ck19} \cite{bock2019multi} and \texttt{b\"{o}ck20} \cite{bock2020deconstruct}. By multi-task training, improvement can be observed on \emph{ACM Mirum} and \emph{GTzan}. Especially for \emph{ACM Mirum}, the Accuracy1 is increased by 2.5\%, achieving the best result among all approaches. Because the two test datasets are both multi-genre datasets, it can be considered that the good performance comes from not only the multi-task learning, but also the beat tracking datasets with rich music genres. As for \emph{GiantSteps}, \texttt{mgan+} performs better than \texttt{b\"{o}ck19} and \texttt{b\"{o}ck20}, but a bit worse than \texttt{mgan}. This is also due to the supplement of data, which affects the dominant position of dance music in training data.

\begin{figure}[tbp]
    \subcaptionbox{\small\label{subfig:multi0}\emph{Samba/Albums-Latin Jam2-14}\quad(0:18-0:24)}{
		\begin{subfigure}{.32\columnwidth}
			\centering
			\includegraphics[width=1in]{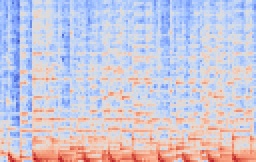}
			\vspace{-1.5mm}
			\caption*{\footnotesize Input Mel-spectrogram}
		\end{subfigure}
		\begin{subfigure}{.32\columnwidth}
			\centering
			\includegraphics[width=1in]{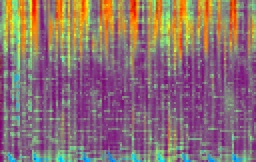}
			\vspace{-1.5mm}
			\caption*{\footnotesize High resolution branch}
		\end{subfigure}
		\begin{subfigure}{.32\columnwidth}
		    \label{fig:multi1}
			\centering
			\includegraphics[width=1in]{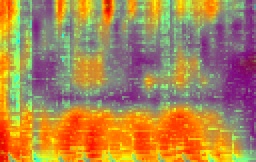}
			\vspace{-1.5mm}
			\caption*{\footnotesize Low resolution branch}
			
		\end{subfigure}
		
		}
	
	\vspace{2.5mm}
	\subcaptionbox{\small \label{subfig:multi1} \emph{Chacha/Albums-Latin Jam-01}\quad(0:06-0:12)}{
	   
		\begin{subfigure}{.32\columnwidth}
			\centering
			\includegraphics[width=1in]{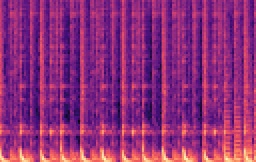}
			\vspace{-1.5mm}
			\caption*{\footnotesize Input Mel-spectrogram}
		\end{subfigure}
		\begin{subfigure}{.32\columnwidth}
			\centering
			\includegraphics[width=1in]{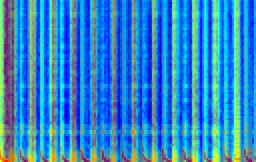}
			\vspace{-1.5mm}
			\caption*{\footnotesize High resolution branch}
		\end{subfigure}
		\begin{subfigure}{.32\columnwidth}
			\centering
			\includegraphics[width=1in]{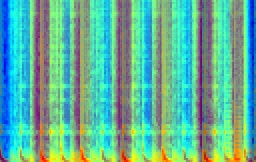}
			\vspace{-1.5mm}
			\caption*{\footnotesize Low resolution branch}
		\end{subfigure}
		
		}
		\caption{Grad-CAM visualizations for layers on different resolution branches.}
		\label{fig:multi}
	\end{figure}

\subsection{Grad-CAM Analysis}
\begin{figure}[tbp]
    \subcaptionbox{\small \label{fig:attention0} \emph{Chacha/Albums-Latino Latino-0}\quad(0:00-0:06)}{
		\begin{subfigure}{.32\columnwidth}
			\centering
			\includegraphics[width=1in]{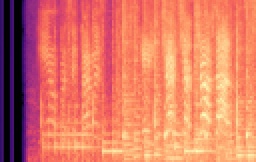}
			\vspace{-1.5mm}
			\caption*{\footnotesize Input Mel-spectrogram}
		\end{subfigure}
		\begin{subfigure}{.32\columnwidth}
			\centering
			\includegraphics[width=1in]{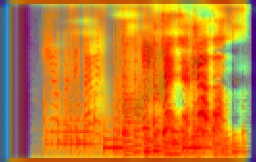}
			\vspace{-1.5mm}
			\caption*{\footnotesize Before attention}
		\end{subfigure}
		\begin{subfigure}{.32\columnwidth}
			\centering
			\includegraphics[width=1in]{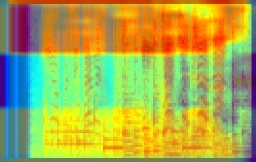}
			\vspace{-1.5mm}
			\caption*{\footnotesize After attention}
		\end{subfigure}
		}
	
	\vspace{2.5mm}
	\subcaptionbox{\small \label{fig:attention1} \emph{Chacha/Albums-Media-103405}\quad(0:12-0:18)}{
		
		\begin{subfigure}{.32\columnwidth}
			\centering
			\includegraphics[width=1in]{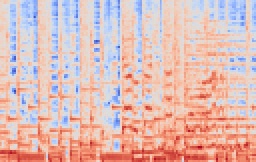}
			\vspace{-1.5mm}
			\caption*{\footnotesize Input Mel-spectrogram}
		\end{subfigure}
		\begin{subfigure}{.32\columnwidth}
			\centering
			\includegraphics[width=1in]{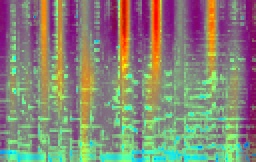}
			\vspace{-1.5mm}
			\caption*{\footnotesize Before attention}
		\end{subfigure}
		\begin{subfigure}{.32\columnwidth}
			\centering
			\includegraphics[width=1in]{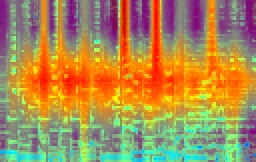}
			\vspace{-1.5mm}
			\caption*{\footnotesize After attention}
		\end{subfigure}
		}
		\caption{Grad-CAM visualizations for layers before and after grouped attention.}
		\label{fig:attention}
	\end{figure}

Gradient-weighted Class Activation Mapping (Grad-CAM) \cite{grad-cam} is a method that can faithfully highlight the important regions in inputs for a CNN-based classification model. It uses the gradient information in back-propagation as weights (grad-weights) to explain the network's decisions. We visualize the activation maps derived by Grad-CAM as shown in Figure \ref{fig:multi} and Figure \ref{fig:attention}. Red indicates the part more important in predicting tempo while blue contributes less.

Figure \ref{fig:multi} shows the activation maps on branches with different resolutions. Their inputs are two audio clips from \emph{Ballroom} dataset. Time duration is marked below the corresponding images, following the audio title set in \emph{italic}. Figure \ref{subfig:multi0} comes from a piece of Samba mainly played by piano and kick drum. The piano in the clip has a higher pitch, played with quarter notes while the kick drum falls on every beat in the bar. It can be observed from the activation maps that the model mainly focuses on short-duration parts of piano in the high-resolution branch, and the kick drum parts with long duration in the low-resolution branch. As for the second example, which is a Cha Cha song, the beat positions can be identified from kick drum in low-frequency part, vocal in middle-frequency part, and claves in high-frequency part. Figure \ref{subfig:multi1} shows that the low-resolution branch considers downbeats to be important, while the high-resolution branch focus on not only downbeats but every other beat in a bar. It can be proved that the multi-scale structure is capable of integrating useful information with different granularities.

We also visualize the activation maps before and after the proposed grouped channel attention to explore the its effect. The results are shown in Figure \ref{fig:attention}. The music excerpt of Figure \ref{fig:attention0} is played with regular claves and double bass, hence the high-frequency part and the low-frequency part contribute more to tempo estimation. The attention branch reweights the feature maps from the trunk branch, giving top and bottom parts higher weights to detect tempo information easier. In contrast, the vocal dominates the 
rhythm information in the song of Figure \ref{fig:attention1}, thus the model gives higher attention to the middle-frequency part after grouped attention. By grouped attention, the network can efficiently find which part would be considered to be important for tempo estimation.

\section{Conclusion}
In this paper, we propose a new CNN-based single-step approach for tempo estimation. We introduce the idea of multi-scale network to construct the architecture of the proposed MGANet. The GAModule with the grouped channel attention is designed to be the key component of the network. Compared with previous work, the proposed approach exhibits good performance on Accuracy1 and outperforms existing models in most cases.


\section{Acknowledgement}
This work was supported by National Key R\&D Program of China (2019YFC1711800), NSFC (61671156).

\bibliography{ISMIR2021_template}

%
%
%
%
%

\end{document}